\documentclass[
reprint, 
superscriptaddress,
 amsmath,amssymb,
 pre,
]{revtex4-1}

\usepackage{graphicx}
\usepackage{dcolumn}
\usepackage{bm}
\usepackage{todonotes}

\usepackage{color}

\newcommand{\diff}[2]{\frac{d#1}{d#2}}
\newcommand{\pdiff}[2]{\frac{\partial #1}{\partial #2}}

\newcommand{\new}{\nonumber\\}
\newcommand{\abs}[1]{\left|#1\right|}

\newcommand{\bvar}{\bm{\varepsilon}}
\newcommand{\bX}{\bm{X}}
\newcommand{\bF}{\bm{F}}
\newcommand{\be}{\bm{e}}
\newcommand{\bY}{\bm{Y}}

\newcommand{\bxi}{\bm{\xi}}
\newcommand{\ave}[1]{\left\langle #1 \right\rangle}

\newcommand{\HH}{\mathcal{H}}

\usepackage{amsmath}	
\begin{document}

\preprint{APS/123-Qed} \title{ Force balance controls the relaxation
time of the gradient descent algorithm in the satisfiable phase}
\author{Sungmin Hwang}
\affiliation{%
LPTMS, Université Paris-Sud 11, UMR 8626 CNRS, Bât. 100, 91405 Orsay
Cedex, France
}%

\author{Harukuni Ikeda}
 \email{harukuni.ikeda@ens.fr}
\affiliation{ Laboratoire de Physique de l'\'Ecole Normale Sup\'erieure,
Universit\'e PSL, CNRS, Sorbonne Universit\'e, Universit\'e de Paris,
75005 Paris, France}


\date{\today}

\begin{abstract}  
We numerically study the relaxation dynamics of the single layer
perceptron with the spherical constraint. This is the simplest model of
neural networks and serves a prototypical mean-field model of both
convex and non-convex optimization problems. The relaxation time of the
gradient descent algorithm rapidly increases near the SAT-UNSAT
transition point. We numerically confirm that the first non-zero
eigenvalue of the Hessian controls the relaxation time. This first
eigenvalue vanishes much faster upon approaching the SAT-UNSAT
transition point than the prediction of Marchenko-Pastur law in random
matrix theory derived under the assumption that the set of unsatisfied
constraints are uncorrelated.  This leads to a non-trivial critical
exponent of the relaxation time in the SAT phase. Using a simple scaling
analysis, we show that the isolation of this first eigenvalue from the
bulk of spectrum is attributed to the force balance at the SAT-UNSAT
transition point.  Finally, we show that the estimated critical exponent
of the relaxation time in the non-convex region agrees very well with
that of frictionless spherical particles, which have been studied in the
context of the jamming transition of granular materials.
\end{abstract}

\pacs{
05.20.-y, 
75.10.Nr, 
87.18Sn, 
}
                             

\maketitle 

\section{Introduction}


Constraint satisfaction problems (CSP) are ubiquitous in physics,
chemistry, and engineering. Since the pioneering paper by Kirkpatrick
\textit{et.~al.}~\cite{kirkpatrick1983optimization}, CSP have been
studied extensively using tools of statistical
mechanics~\cite{nishimori2001statistical,mezard2009information}.
Although numerous studies have been done for CSP involving discrete
degrees of freedom such as
K-SAT~\cite{nishimori2001statistical,mezard2009information}, the study
of problems with continuous degrees of freedom is still in its
infancy~\cite{franz2017universality}.

The standard approach of statistical mechanics is to first consider
solvable mean-field models. Among several models of CSP with the
continuous degrees of freedom, the perceptrons are probably most
popular~\cite{rosenblatt1958perceptron}. They are the
simplest models of neural networks working as linear classifier
of the given data set. If the size of input dataset is small, the system
is in a satisfiable (SAT) phase where one can find neural weights that
can perfectly classify the entire data. Contrary, if the size is too
large, the system lies in the unsatisfiable (UNSAT) phase where no such
solutions exist. In the thermodynamic limit, the SAT-UNSAT transition
becomes a sharp phase transition at which several physical quantities
exhibit singular behaviors~\cite{gardner1988optimal}.

The static equilibrium properties of the perceptron are now
well-understood due to sophisticated mean-field theories such as the
replica
method~\cite{gardner1988optimal,franz2016simplest,franz2017universality}.
However, their understanding of dynamics is still far from
complete~\cite{agoritsas2018out}. In this paper, we study the quench
dynamics of the perceptron through an extensive numerical simulation.
We use the gradient descent dynamics (GDD), which is the most basic
algorithm to optimize the cost function of neural networks including the
perceptrons~\cite{lecun2015deep,nielsen2015neural}. In particular, it is
important to understand the dynamics in the SAT phase ({\it i.e.},
overparameterized phase) where the number of model parameters is larger
than the number of input data. Many modern neural networks are trained
in such regions~\cite{nielsen2015neural}, because overparameterized
models can relax faster and avoid to get stuck in a bad local minima
where the cost function has a higher
value~\cite{soudry2016no,lipton2016stuck,cooper2018loss,Advani2017,Draxler2018,Geiger2019a}.


Another motivation to study the perceptron with GDD is closely related
to the dynamics of granular materials. The viscosity of driven granular
particles diverges at a certain density, which is the so-called jamming
transition~\cite{olsson2007critical}. It has been well-established that
the jamming transition of spherical frictionless particles belongs to
the same universality class of the SAT-UNSAT transition of the
perceptron in the large dimensional
limit~\cite{franz2016simplest,franz2017universality}, whereas in finite
dimensions, it is far less clear due to to the existence of the
nontrivial finite-dimensional features such as the localized
modes~\cite{lerner2013low,charbonneau2015jamming} and spatial
fluctuation~\cite{hikeda2019universal}.

Interestingly, a
recent numerical simulation of spherical particles reveals that the
relaxation time of particle systems driven by GDD is proportional to the
shear viscosity of the shear driven system near the jamming transition
point~\cite{ikeda2019universal}. This suggests that the perceptron
driven by GDD would be the simplest model to study the dynamics of the
jamming transition.

In this work, by combining an extensive numerical simulation and scaling
theory developed in Refs.~\cite{lerner2012unified,lerner2012toward}, we
show that the relaxation time in the SAT phase is controlled by the
\textit{unbalanced} force, which is the net force divided by the square
root of the energy. By construction, the unbalanced force vanishes in
the UNSAT phase, which leads to the divergence of the relaxation time
when the system approaches to the SAT-UNSAT transition point from the
SAT phase. Furthermore, interestingly, we find that the critical
exponent obtained by our numerical simulation agrees very well with one
of the theoretical prediction for the shear viscosity of spherical
particles~\cite{lerner2012toward}.

The paper is organized as follows.  In Sec.~\ref{161400_12Aug19}, we
introduce the model. In Sec.~\ref{163936_12Aug19}, we show our numerical
result for the relaxation dynamics. In Sec.~\ref{180609_23Sep19}, we
discuss that an isolated eigenmode appears near the transition point,
and this isolated mode controls the relaxation time.  In
Sec.~\ref{180639_23Sep19}, we discuss the scaling theory of the isolated
mode. Finally, in Sec.~\ref{164328_12Aug19}, we summarize and conclude
the work.

\section{Setting}
\label{161400_12Aug19}

\subsection{Model}
In this work, we consider the generalized perceptron model investigated
by Franz {\it et al.}~\cite{franz2016simplest}. In this section, we
describe the detailed definition of the model in the context of the
constraint satisfaction problem.

The perceptron model was originally introduced by
Rosenblatt~\cite{rosenblatt1958perceptron}. The aim of the perceptron is
to correctly classify the input data. More precisely, one wants to find
out the state variable $\bX=\{X_1,\cdots, X_N\}$ such that
\begin{align}
 {\rm sgn}\left[\frac{1}{\sqrt{N}}\bX\cdot\tilde{\bxi}^\mu\right] =y_\mu,\label{165754_31Aug19}
\end{align}
for all $M$ input-output associations of $\tilde{\bxi}^\mu =
\{\tilde{\xi}_1^\mu,\cdots,\tilde{\xi}_N^\mu\}$ and $y_\mu \in
\{-1,1\}$ where $\mu$ is an index running from $1$ to $M$.  Since these
constraints are scale-independent, it is natural to introduce a
regularization condition
\begin{align}
\bX\cdot\bX = N.\label{165743_22Jul19} 
\end{align}
to prevent an overflow through the dynamics.  Additionally, let us
consider the case where $\tilde{\xi}_i^\mu$ is a Gaussian random
variable with zero mean and unit variance.

The classification problem above can be recast into a constraint
satisfaction problem with the following constraints
\begin{align}
h_\mu  = \frac{y_\mu}{\sqrt{N}}\bX\cdot\tilde{\bxi}^\mu =
 \frac{1}{\sqrt{N}}\bX\cdot\bxi^\mu \geq 0,\label{172308_31Aug19}
\end{align}
where we have introduced new random variables $\bxi^\mu = y_\mu
\tilde{\bxi}^\mu$, which has the same distribution of the original
one. 
A conventional approach to solve this problem is to translate it into an optimization problem with a corresponding
cost function
\begin{align}
 H &= \sum_{\mu=1}^M \frac{h_\mu^2}{2}\theta(-h_\mu),\label{201816_31Aug19}
\end{align}
where $\theta(x)$ denotes the Heaviside step function. This cost
function is designed in such a way that it vanishes $H=0$ if and only if $\bX$ satisfies all the
constraints Eq.~(\ref{172308_31Aug19}).

The typical case performance of the perceptron can be studied by
calculating the typical value of $H$ at zero temperature, which is
tantamount to studying the ground state energy of the model where the
interaction among state variables $X_i$'s are given by the Hamiltonian
$H$. This detailed thermodynamic study uncovers a sharp phase transition
in the thermodynamics limit from a satisfiable (SAT) phase, where one
can find $\bX$ such that $H=0$, to an unsatisfiable (UNSAT) phase, where
there are no such configurations and thus
$H>0$~\cite{nishimori2001statistical}.

The cost function of the original perceptron model can be shown to be
convex, and thus to form a single cluster of solutions.  However, in
many realistic problems, such as the state of the art multilayer neural
networks used in machine learning algorithms, the corresponding
optimization problems are not necessarily convex and the cost function
can have multiple minima~\cite{nielsen2015neural}. To investigate the
effect of non-convexity, Franz {\it et al.}~\cite{franz2016simplest}
introduced a variant of the standard perceptron with the following
modified constraints:
\begin{align}
 h_\mu = \frac{1}{\sqrt{N}}\bX\cdot\bxi^\mu - \sigma \geq 0 ,\ \mu = 1,\cdots, M,
\end{align}
where $\sigma$ is referred to as the bias. The original problem
corresponds to $\sigma=0$. One can define the cost function as
Eq.~(\ref{201816_31Aug19}), and calculate the phase diagram as a
function of $\sigma$ and $\alpha=M/N$ by using the replica method.  As
in the case of the standard perceptron, the model exhibits the SAT-UNSAT
transition at $\alpha=\alpha_c$ at which $H$ begins to have a non-zero
value~\cite{franz2016simplest}. When $\sigma\geq 0$, the cost function
has a single minimum, and thus the optimization problem is convex as in
the case of the standard perceptron~\cite{gardner1988optimal}. On the
contrary, as soon as $\sigma<0$, the cost function can form multiple
minima depending on the choice of input-output associations. In
particular, it is known that near $\alpha_c$ the typical realization of
this problem is always
non-convex~\cite{franz2016simplest,franz2017universality}.

The static critical behavior of the perceptron in the non-convex region
$\sigma<0$ has been fully investigated using the replica method.  At the
SAT-UNSAT transition point, the theory predicts that (i) the system
becomes isostatic at the SAT-UNSAT transition point, meaning that the
contact number is the same of that of the number of degrees of
freedom~\cite{franz2017universality}, (ii) the two point correlation
function exhibits power law scaling, for instance, the two point force
distribution has a pseudo gap $P(f)\sim f^\theta$ with $\theta=0.423$
for small force $f$~\cite{franz2017universality}, and (iii) the
eigenvalue distribution is gapless in the UNSAT
phase~\cite{franz2015universal}.

\subsection{Dynamics}
We consider the simple GDD:
\begin{align}
 \diff{\bX(t)}{t} = -P(t)\cdot\nabla H,\label{172119_22Jul19}
\end{align}
where $\nabla_i = \partial/\partial X_i$, and 
\begin{align}
 P_{ij}(t) &= \delta_{ij} - \frac{1}{N}X_i(t)X_j(t)\label{195234_5Sep19}
\end{align}
denotes the projection operator onto a hypersphere defined by
Eq.~(\ref{165743_22Jul19}). Using Eq.~(\ref{172119_22Jul19}), one can
show that
\begin{align}
\bX(t)\cdot\dot{\bX}(t)=0,
\end{align}
suggesting that the constraint Eq.~(\ref{165743_22Jul19}) is 
automatically satisfied if $\bX(0)\cdot\bX(0)=N$. For the numerical
integration, we have to discretize Eq.~(\ref{172119_22Jul19}) without
violating Eq.~(\ref{165743_22Jul19}). For this purpose, we consider the
following discretized dynamics:
\begin{align} 
 \bY(t+\Delta t) &=
 \bX(t) -\Delta t \nabla H   ,\new
 \bX(t+\Delta t) &=
 \sqrt{N}\frac{\bY(t+\Delta t)}{\sqrt{\bY(t+\Delta t)\cdot\bY(t+\Delta t)}},\label{171032_22Jul19}
\end{align}
where $\Delta t$ denotes the time step. One can show that
Eq.~(\ref{171032_22Jul19}) agrees with Eq.~(\ref{172119_22Jul19}) up
to the first order of $\Delta t$.

\subsection{Details of numerics}
For the initial condition $\bX(0)$, we generate a uniform random
configuration on the $N$ dimensional hypersphere so that
$\bX(0)\cdot\bX(0) = N$.  Starting from this configuration, we evolve
the system by applying Eq.~(\ref{171032_22Jul19}) iteratively.  We
define the time as $t=\Delta t N_{\rm step}$ where $N_{\rm step}$
denotes the number of the iteration.  We stop the iteration when
\begin{align} 
f_p \equiv \sqrt{\frac{1}{N}\sum_{i=1}(P\nabla H)_i^2}
 < 10^{-10}.\label{081728_4Sep19}
\end{align}
We use $\Delta t = 0.1$ and $N=256$ unless otherwise noted. Hereafter we
mostly show numerical results for $\sigma=0.5$, where the cost
function is convex, and $\sigma=-0.5$, where the cost function is
non-convex.

\section{Relaxation}
\label{163936_12Aug19}

\subsection{Time evolution of physical quantities}

\begin{figure}[t]
\includegraphics[width=9cm]{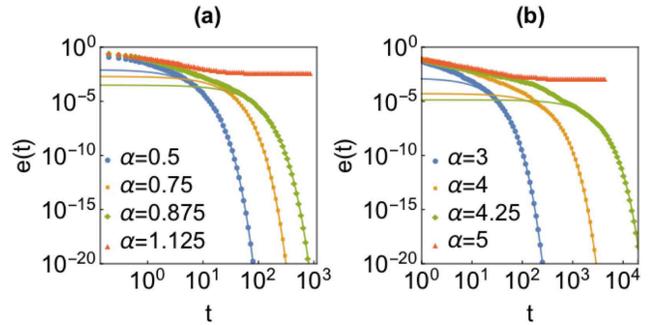} \caption{ Time evolution of the
 energy as a function of time (a) in a convex regime $\sigma=0.5$ and
 (b) in a non-convex regime $\sigma=-0.5$. The markers denote the
 numerical results of single trajectory for each parameter. The solid
 lines denote the exponential fit.  } \label{122402_1Sep19}
\end{figure}
First, we report the time evolution of several physical quantities. In
Fig.~\ref{122402_1Sep19} (a), we show the time dependence of the cost
function per degree of freedom $e(t) = H(t)/N$ for $\sigma=0.5$ where
the static replica calculation predicts that the cost function is
convex~\cite{franz2017universality}. For small $\alpha$, $e(t)$
decreases monotonically to zero, see the data for $\alpha=0.5$, $0.75$,
and $0.875$ in Fig.~\ref{122402_1Sep19} (a). This indicates that the
system lies in a SAT phase. The late time behavior of $e(t)$ can be well
fitted by an exponential function (see the solid lines in
Fig.~\ref{122402_1Sep19} (a)).  On the contrary, for the larger values
of $\alpha$, $e(t)$ does not decay to zero in the long time limit,
indicating that the system lies in an UNSAT phase (see the data for
$\alpha=1.125$ in Fig.~\ref{122402_1Sep19} (a)). In
Fig.~\ref{122402_1Sep19} (b), we show $e(t)$ for $\sigma=-0.5$ where the
cost function is non-convex~\cite{franz2017universality}. Despite this
difference, the relaxation of $e(t)$ is quite similar to that of the
convex case ($\sigma=0.5$). Namely, $e(t)$ exhibits an exponential decay
for small $\alpha$, (see the data for $\alpha=3$, $4$, and $4.25$),
while it converges to a finite value for larger $\alpha$'s, (see the
data for $\alpha=5$). Further studies, such as the investigation of the
aging dynamics, are necessary to clarify the qualitative difference of
the relaxation dynamics between the convex and non-convex problems. We
leave it for future work.

\begin{figure}[t]
\includegraphics[width=9cm]{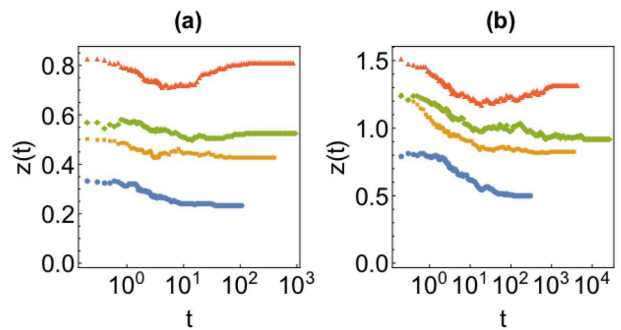} \caption{ Time dependence of the
 contact number. The markers denote the numerical results of single
 trajectory for each parameter. The values of $\alpha$ are the same as
 Fig.~\ref{122402_1Sep19}. (a) The results for the convex problem
 $\sigma=0.5$.  (b) The result for the non-convex problem $\sigma=-0.5$.
 } \label{115124_2Sep19}
\end{figure}

The other important quantity is the fraction of unsatisfied constraints:
\begin{align}
 z(t) &= \frac{1}{N}\sum_{\mu=1}^M \theta(-h_\mu).\label{120225_2Sep19}
\end{align}
Following the analogy of the jamming of particle systems, 
we shall call $z(t)$ the "contact number". In Fig.~\ref{115124_2Sep19} (a), we
show the time evolution of $z(t)$ for $\sigma=0.5$ and the same values
of $\alpha$ as Fig.~\ref{122402_1Sep19} (a). $z(t)$ converges to a
finite value in the long time limit:
\begin{align}
z\equiv \lim_{t\to\infty}z(t).\label{173809_8Oct19}
\end{align}
$z$ tends to smoothly increase with $\alpha$. It may be a little counter
intuitive that $z$ has a finite value even in the SAT phase where
$e\equiv \lim_{t\to\infty}e(t)=0$. However, this is a natural
consequence of GDD Eq.~(\ref{172119_22Jul19})
and definition of $z(t)$ Eq.~(\ref{120225_2Sep19}). Since the dynamics
does not involve inertia, some contacts converge to $h_\mu\to
0^{-}$, implying that $\theta(-h_\mu)=1$ even in the long time
limit~\cite{ikeda2019universal}.

\subsection{Physical quantities at the stationary state}

Next, we shall study $e(t)$ and $z(t)$ in the long time limit,
$e\equiv \lim_{t\to\infty}e(t)$ and $z\equiv \lim_{t\to\infty}z(t)$.  To
obtain the stationary state configuration, we run numerical
simulations for various values of $\alpha$ and initial conditions until
Eq.~(\ref{081728_4Sep19}) is satisfied. Then, we calculate the energy
$e(t)$ and the contact number $z(t)$ at the stationary state.

\begin{figure}[t]
\includegraphics[width=9cm]{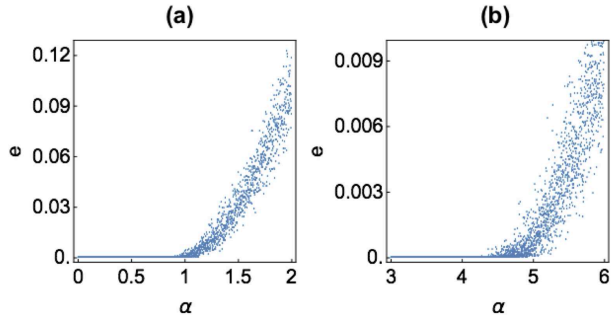} \caption{ $\alpha$ dependence of
the energy per degree of freedom at the stationary state. The markers denote the
numerical results measured for each individual configuration.  (a)
Numerical results in the convex phase.  (b) Numerical results in the
non-convex phase.  } \label{123200_3Sep19}
\end{figure}
In Fig.~\ref{123200_3Sep19} (a), we show the stationary state energy $e$
for the convex case $\sigma=0.5$. For small $\alpha$, the zero energy
$e=0$ suggests that the system lies in a SAT phase. The energy $e$
begins to have a non-zero value at $\alpha=\alpha_c\approx 1$, which is
the signature of the SAT-UNSAT transition. The transition point well
agrees with the theoretical prediction $\alpha_c =
0.961$~\cite{franz2017universality}.  In Fig.~\ref{123200_3Sep19} (b),
we show the numerical result for the non-convex case $\sigma=-0.5$. The
SAT-UNSAT transition takes place at $\alpha_c\approx 4.5$.
This is close to a theoretical prediction
$\alpha_c=4.77$ in Ref.~\cite{franz2017universality}.
While this prediction is made under the assumption
that the problem remains convex near jamming, it still provides a good agreement
in the parameter range where the simulation is performed.

\begin{figure}[t]
\includegraphics[width=9cm]{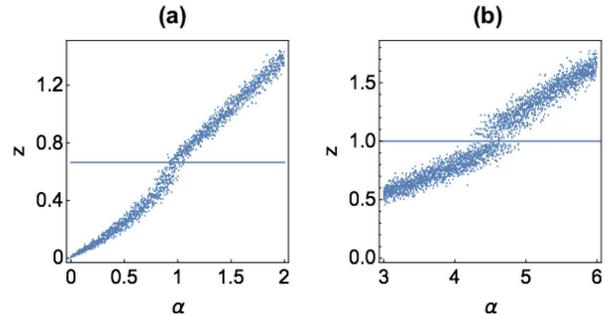} \caption{ $\alpha$ dependence of
the contact number per degree of freedom at the stationary state. The markers
denote the numerical results measured for each individual configuration.
The solid line denotes the theoretical prediction of $z$ at the
SAT-UNSAT transition point.  (a) Numerical results in the convex phase.
(b) Numerical results in the non-convex phase.  } \label{064548_4Sep19}
\end{figure}
In Fig.~\ref{064548_4Sep19} (a), we show the contact number of the
stationary state $z$ for the case of the convex problem ($\sigma=0.5$).
$z$ tends to increase in $\alpha$ and develops a non-analytic point
exactly at the SAT-UNSAT transition. The numerical results of $z$ at
$\alpha_c$ well agrees with the theoretical prediction
$z(\alpha_c)=0.665$, i.e., the horizontal lines originally computed in
~\cite{franz2017universality}.  In Fig.~\ref{064548_4Sep19} (b), we show
$z$ for the case of the non-convex problem $(\sigma=-0.5)$. The theory
predicts that for $\sigma<0$, the system becomes isostatic at the
transition point,
 $z(\alpha_c)=1$~\cite{franz2017universality}. The numerical
result agrees well with this prediction, see the horizontal
line in Fig.~\ref{064548_4Sep19} (b).

  \subsection{Relaxation time}

Finally, we discuss the $\alpha$ dependence of the relaxation time.  We
define the relaxation time $\tau$ as the time when the system
first satisfies the stationary state condition Eq.~(\ref{081728_4Sep19}).

\begin{figure}[t]
\includegraphics[width=9cm]{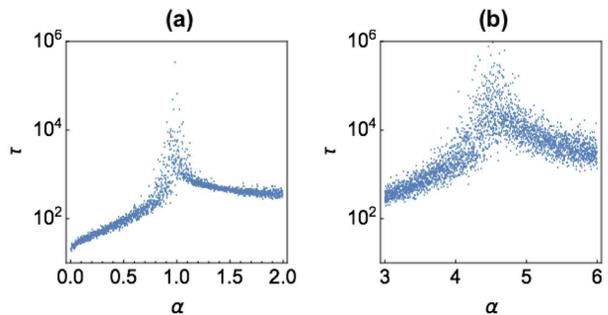} \caption{ $\alpha$
 dependence of the relaxation time. The markers denote the numerical
 results measured for each individual configuration. (a) The result in
 the convex phase.  (b) The result in the non-convex phase. }
 \label{093955_5Sep19}
\end{figure}
In Fig.~\ref{093955_5Sep19}, we show the $\alpha$ dependence of
$\tau$ for both a convex regime ($\sigma=0.5$) and a non-convex regime ($\sigma = -0.5$).
The relaxation time $\tau$ exhibits a sharp peak at the SAT-UNSAT transition point (a)
$\alpha_c\approx 1$ and (b) $\alpha_c\approx 4.5$. 
These results prove that the SAT-UNSAT transition is a critical
phenomenon accompanied by the divergence of the relaxation time. Below,
we show that this divergence is a consequence of vanishing first nonzero
eigenvalue of the Hessian of the cost function.

\section{Eigenmodes of Hessian}
\label{180609_23Sep19}

We here investigate the eigenvalues of the Hessian constructed from a
second-order approximation of the cost function evaluated at the
stationary point.  In a SAT phase, the stationary point is formed at a
boundary of solution space due to a lack of inertia.  Thus, one can
naturally expect that there exist many zero modes along the directions
towards islands of solutions while there are also non-zero modes coming
from the contributions of infinitesimally unsatisfied patterns at the
boundary.  If these patterns are statistically uncorrelated, one can
immediately show that the spectrum of such Hessian follows a
Marchenko-Pastur law~\cite{franz2015universal}.

Strikingly, one of our main findings is to show that the first non-zero
eigenmode is an outlier when compared against this null model.
Specifically, we found that the eigenvalue is statistically much smaller
than the bulk spectrum which cannot be explained by a usual Tracy-Widom
distribution.  This implies that our dynamics chooses the set of
unsatisfied patterns in such a way that they form a non-trivial
correlation.  Because the dynamics should be well approximated by a
corresponding Hessian dynamics at least near the boundary, we can
conclude that the dynamics is significantly slower than that of
relaxation dynamics of random patterns.

\subsection{Derivation of the Hessian at the stationary state in the SAT phase}
We expand the cost function around the stationary state $\bX_*$ as
follows
\begin{align}
\delta H &= H(\bX_*+P\bvar) -H(\bX_*) \new
 &\approx   P\bvar\cdot\nabla H +
 \frac{1}{2}\left[
 P\bvar\cdot\nabla \left(P\bvar\cdot\nabla H\right)
 \right]\new
 &= \bvar\cdot P \cdot \nabla H
 + \frac{1}{2}\bvar\cdot M \cdot \bvar\label{141153_24Jul19}
\end{align}
where $P$ denotes the projection operator defined by
Eq.~(\ref{195234_5Sep19}). One can always eliminate the anti-symmetric
part of $M$ in Eq.~(\ref{141153_24Jul19}) and express it as a symmetric
matrix
\begin{align}
 M_{ij} &= \frac{1}{2}\sum_{n,m=1}^N\Bigg{[}
 P_{in}\pdiff{}{X_n}\left(P_{jm}\pdiff{H}{X_m}\right)\\
& + P_{jn}\pdiff{}{X_n}\left(P_{im}\pdiff{H}{X_m}\right)
\Bigg{]}\new
 &= \left(P\cdot\nabla^2 H\cdot P\right)_{ij} + \zeta P_{ij} 
 -\frac{(P\nabla H)_i X_j + X_i (P\nabla H)_j}{2N},
\end{align}
where we have introduced an auxiliary variable
\begin{align}
\zeta &= -\frac{1}{N}\bX\cdot\nabla H
 = -\frac{1}{N}\sum_{\mu=1}^M\theta(-h_\mu)\left(h_\mu^2 + \sigma h_\mu\right).
\end{align}
At the stationary state, we have $(P\nabla H(\bX_*))_i = \sum_{n=1}^N
P_{in}\nabla_n H =0$.  Also, in the SAT phase, $h_\mu \to 0$ which leads
to $\zeta = 0$. Under such conditions, the Hessian matrix $\HH$ can be
simply expressed as
 \begin{align}
 \HH_{ij}  &= \frac{1}{N}\left(P\cdot \nabla^2 H \cdot P\right)_{ij}
=   \frac{1}{N}\sum_{\mu=1}^M\theta(-h_\mu)
  \left(P\bxi^\mu\right)_i\left(P\bxi^\mu\right)_j.\label{121817_1Aug19}
 \end{align}
Note that this expression is different from the one studied in the
equilibrium dynamics in Ref.~\cite{altieri2016jamming} where authors
considered the Hessian of the free energy.

\subsection{Zero modes}

There are $N(1-z)$ number of linearly independent vectors $\be_l$,
$l=1,\cdots, N(1-z)$ that satisfy $\be_l\cdot P\bxi^{\alpha}=0$ for
$\alpha=1,\cdots, Nz$, where $z$ denotes the number of contacts
normalized by $N$ at the stationary state given by
Eqs.~(\ref{120225_2Sep19}) and (\ref{173809_8Oct19}), and $\bxi^\alpha$
denotes the contact that satisfies $h_{\alpha}\leq 0$. From
Eq.~(\ref{121817_1Aug19}), it follows that
\begin{align}
 \HH \be_l = 0,
\end{align}
meaning that $\be_l$ is a zero eigenvector of $\HH$. Since the system
does not evolve along the direction of the zero modes, hereafter we
neglect the zero modes.

\subsection{Isolated eigenmode}

\begin{figure}[t]
\includegraphics[width=9cm]{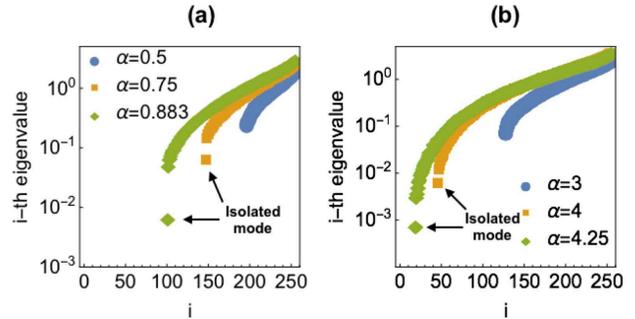} \caption{ Examples of spectrum of
Hessian matrices shown in rank plots. The markers denote the numerical
results for a single realization for each parameter.
 Zero modes
are excluded from the plots. (a) The results for the convex problem
($\sigma=0.5$).  (b) The result for the non-convex problem
($\sigma=-0.5$).}  \label{151720_9Sep19}
\end{figure}
In Fig.~\ref{151720_9Sep19} (a) and (b), we show the typical behavior of
the eigenvalues at the stationary state in the SAT phase for the case of
the convex problem ($\sigma=0.5$) and non-convex problem
($\sigma=-0.5$), respectively. In the SAT phase, there are $N(1-z)$
number of zero modes (not shown). As $\alpha$ approaches $\alpha_c$, the
first nonzero eigenvalue $\lambda_1$ decreases much faster than the
other eigenvalues, suggesting that $\lambda_1$ is the isolated
eigenvalue near $\alpha_c$ for both convex and non-convex problems.

\subsection{Eigenvalues and relaxation time in the SAT phase}

\begin{figure}[t]
\includegraphics[width=9cm]{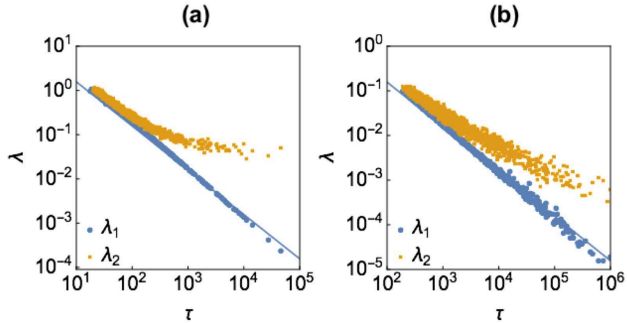} \caption{ Scatter plots of the
first/second non-zero eigenvalues and the relaxation time. The markers
denote the numerical results measured for each individual
configuration. The solid line indicates $\lambda\propto \tau^{-1}$.  (a)
The results for the convex problem ($\sigma=0.5$).  (b) The result for
the non-convex problem ($\sigma=-0.5$).  } \label{173940_15Sep19}
\end{figure}
In Fig.~\ref{173940_15Sep19} (a) and (b), we show the scatter plots of
the first and second eigenvalues, $\lambda_{1}$ and $\lambda_{2}$, against 
the relaxation time $\tau$ in the SAT phase. One can clearly see that $\tau$ is
inversely proportional to $\lambda_{1}$:
\begin{align}
 \tau \sim \lambda_{1}^{-1}.
\end{align}
This is a direct evidence of the fact that $\lambda_{1}$ controls 
GDD near the SAT-UNSAT transition point. The second smallest eigenvalue
$\lambda_{2}$ behaves similarly to $\lambda_{1}$ for small $\tau$, while
it starts to deviate from $\lambda_1$ as $\tau$ increases. This implies
that the separation of $\lambda_{1}$ and $\lambda_2$ becomes more
pronounced as the system approaches the transition point. This is
consistent with the results shown in Fig.~\ref{151720_9Sep19}, where the
first eigenvalue is isolated near the transition point.

\subsection{Scaling of the first eigenvalue}

Recent numerical studies of a particle system reveal that the relaxation
time $\tau$ of the quench system is proportional to the shear
viscosity $\eta$ near the jamming transition point, if one plots both
quantities as a function of the contact
number~\cite{ikeda2019universal}. This motivates us to study the scaling
of $\tau$ of the perceptron for $\sigma<0$ where the model belongs to
the same universality class of spherical particles in the large
dimensional limit~\cite{franz2016simplest}. As discussed in the previous
section, $\tau$ is inversely proportional to $\lambda_{1}$.  Therefore,
instead of $\tau$, we here calculate $\lambda_{1}$ as a function of the
contact number $z$. We perform extensive numerical simulations for
various initial configurations and for different values of $\alpha$.
Obviously, each different setting will find a different value of $z$ in
the stationary limit.  Thus, we calculate the mean value of
$\lambda_{1}$ averaged over the samples with the same value of $z$.  We
collected at least $10$ samples for each $z$.

\begin{figure}[t]
\includegraphics[width=8cm]{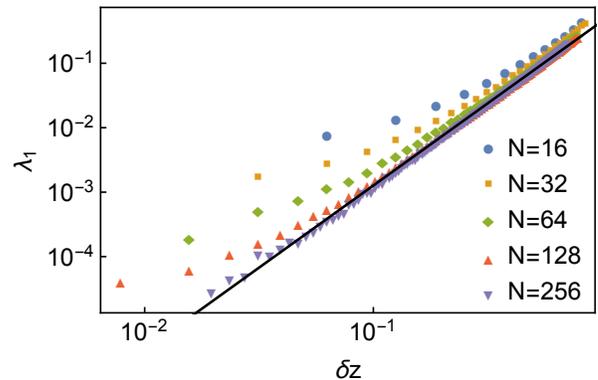} \caption{ Scaling of
 $\lambda_{1}$ for $\sigma=-0.5$.  The markers denote the numerical
 results of the average value of $\lambda_1$, while the solid line
 denotes the prediction of the finite size scaling $\lambda_{1}\sim
 \delta z^{2.55}$. } \label{170340_18Sep19}
\end{figure}
In Fig.~\ref{170340_18Sep19}, we show our numerical results of
$\lambda_{1}$ as a function of the deficit contact number
\begin{align}
 \delta z = 1-z.
\end{align}
For the non-convex region ($\sigma<0$), the perceptron becomes isostatic
$\delta z = 0$ at the transition point~\cite{franz2017universality}. We
find that $\lambda_{1}$ exhibits power law scaling for the
intermediate value of $\delta z$. For very small $\delta z$, however,
$\lambda_{1}$ deviates from the power law and converges to a finite
value. The power law region persists longer as $\delta z \to 0$ for larger
$N$, suggesting that the deviation from the power law is a finite size
effect.

In order to determine the critical exponent precisely, we perform a
finite size scaling analysis. Following the scaling argument above the
jamming transition point~\cite{yan2016variational}, we assume that
$\lambda_{1}\sim \delta z^\beta$ for $\delta z \gg 1/N$, while
$\lambda_{1}$ converges to a finite value for $\delta z \sim 1/N$. This
assumption leads to the following scaling function for finite $N$
systems:
\begin{align}
 \lambda_{1}(N,\delta z) \sim N^{-\beta}f_1(N\delta z),\label{174654_20Sep19}
\end{align}
where $f_1(x)\sim x^\beta$ for $x\gg 1$, and $f_1(x)\sim x^0$ for $x\sim
1$. 
\begin{figure}[t]
\includegraphics[width=9cm]{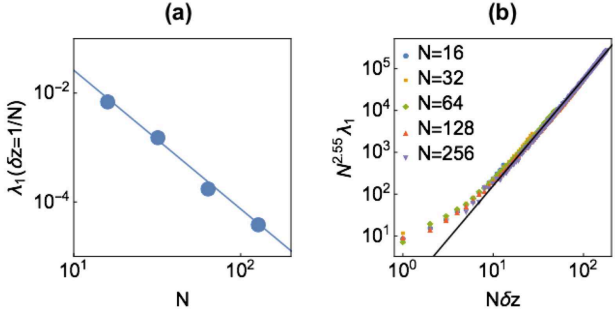} \caption{ Finite $N$
 scaling of $\lambda_{1}$ for $\sigma=-0.5$.  (a) $N$ dependence of
 $\lambda_{1}$ for $\delta z =1/N$. The markers denote the numerical
 result, while the solid line denotes the result of the power law fit
 $\lambda_{1}\sim N^{-2.55}$.  (b) Scaling plot of the same data as
 Fig.~\ref{170340_18Sep19}.}
					       \label{173246_20Sep19}
\end{figure}
In Fig.~\ref{173246_20Sep19} (a), we show the $N$ dependence of
$\lambda_{1}$ for $\delta z=1/N$. As expected from
Eq.~(\ref{174654_20Sep19}), the data are well fitted by a power law
$\lambda_{1}\sim N^{-\beta}$ with a critical exponent
\begin{align}
 \beta=2.55\pm 0.15.\label{174716_20Sep19}
\end{align}
In Fig.~\ref{173246_20Sep19} (b), we show a scaling plot predicted by
Eqs.~(\ref{174654_20Sep19}) and (\ref{174716_20Sep19}). The excellent
collapse of the data for different $N$ strongly supports our scaling
analysis.

\section{Scaling theory}
\label{180639_23Sep19} In this section, we try to identify the origin
of the isolated mode and derive the analytic expression of the
dynamical critical exponent $\beta$.

\subsection{Unbalanced force controls the isolated mode}

\begin{figure}[t]
\includegraphics[width=8cm]{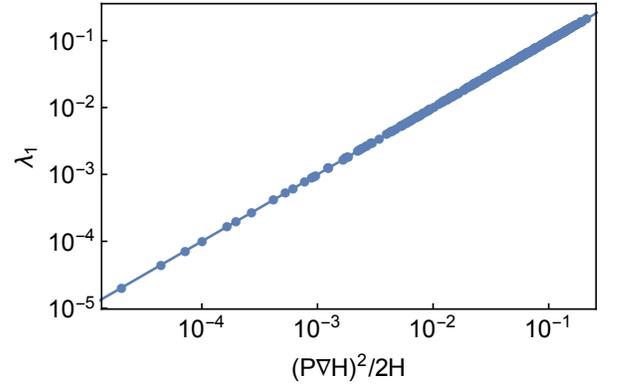} \caption{Scatter plot of
 $\lambda_{1}$ and $(P\nabla H)^2/(2H)$ for $\sigma=-0.5$.  The markers
 denote the numerical results measured for each individual
 configuration. The solid line denotes the theoretical prediction
 $\lambda_1 = (P\nabla H)^2/2H$.}  \label{122557_21Sep19}
\end{figure}

Here we discuss that the force balance at the SAT-UNSAT transition point
leads to the vanishing behavior of $\lambda_{1}$. For this purpose, we
consider the linearized equation around the stationary state $\delta
\dot{\bX}(t) \approx -\HH \delta\bX(t)$, where $\delta
\bX(t)=\bX(t)-\bX_*$, and $\bX_*$ denotes the configuration at the
stationary state. In the long time limit, $\delta\bX(t)$ converges to
the eigenvector of the first non-zero eigenvalue $\lambda_1$, therefore
we have $\delta\dot{\bX}(t)\sim -\lambda_{1}\delta\bX(t)$, which leads
to
\begin{align}
 |\delta\bX(t)| \sim e^{-\lambda_{1}t}.
\end{align}
Similarly, the cost function is $H(t)\sim |\delta\bX(t)^2| \sim
e^{-2\lambda_{1}t}$, which allows us to express $\lambda_{1}$ as
\begin{align}
 \lambda_{1} &=-\frac{1}{2}\lim_{t\to \infty}\frac{\dot{H}(t)}{H(t)}
= 
 \left.\frac{\sum_{i}(P\nabla H)_i^2}{2H}\right|_{\bX=\bX_*}\new
 &\equiv \frac{1}{N}\sum_{i=1}^NF_i^2.\label{123241_21Sep19}
\end{align}
Here we have introduced the {\it unbalanced force} as
\begin{align}
 F_i &= \sum_{\mu=1}^M\theta(f_\mu)\frac{1}{\sqrt{N}}(P\bxi^\mu)_i f_\mu, &
 f_\mu &= -\frac{h_\mu}{\sqrt{\ave{h^2}}}.
\end{align}
$\bF = \{F_1,\cdots, F_N\}$ is the eigenvector of $\lambda_1$, because
$\bF\propto P\nabla H \propto \HH\delta\bX$, and $\delta\bX$ converges
to the eigenvector of $\lambda_1$.  In Fig.~\ref{122557_21Sep19}, we
numerically confirm the validity of Eq.~(\ref{123241_21Sep19}).

 In the UNSAT phase, $F_i=0$ because $H>0$ and $P\nabla H=0$ at the
stationary state. On the contrary, in the SAT phase, $F_i$ can have a
finite value because both $H$ and $P\nabla H$ vanish at the stationary
state. From the continuity of $F_i$, it follows that $F_i=0$ at the
SAT-UNSAT transition point, which leads to $\lambda_{1}=0$ and the
divergence of the relaxation time $\tau$. It is worth noting that the
above scenario, where the force balance controls the slow dynamics near
the transition point, holds not only for the perceptron but also for
more general models driven by GDD both in the convex and non-convex
phases.

\subsection{Variational argument}
\label{164234_12Aug19}

Now we derive the scaling of $\lambda_{1}$ using the assumption of the
marginal stability~\cite{muller2015marginal}.  When the quench rate is
increases, the system arrives at less stable state. In particular, for
GDD, which corresponds to the infinitely fast quench, the system
would reach the most unstable configuration for the given
constraints. More concretely, for our model, among possible
configurations with fixed $z$, the one with the smallest $\lambda_{1}$
would be realized:
\begin{align}
 \lambda_{1} = \min_{{\rm fixed}\ z} \frac{1}{N}\sum_{i=1}^N F_i^2.\label{115457_22Sep19}
\end{align}
Following Ref.~\cite{lerner2012toward}, we shall construct the
configuration satisfying Eq.~(\ref{115457_22Sep19}) by removing the
contacts from the isostatic configuration where $\delta z = 0$ and
$\lambda_{1}=0$.  Removing the contacts would break the force balance,
leading to $F_i>0$ and $\lambda_{1}>0$. In order to minimize
$\lambda_{1}$, one should minimize the perturbation from the isostatic
configuration. This would be possible by removing the weakest contacts
that have the smallest values of $f_\mu$. The typical force scale of the
weakest contacts is
\begin{align}
 f_{\rm av} \equiv \frac{\int_0^{f_*}P(f)f df}{\int_0^{f_*}P(f) df} \sim f_{*},
\end{align}
where $P(f)\sim f^{\theta}$ with $\theta=0.423$ denotes the force
distribution at
jamming~\cite{franz2017universality,charbonneau2014fractal}, and the
upper bound $f_*$ is calculated by using the extreme value statistics as
follows
\begin{align}
\int_0^{f_*}P(f)df
 \sim \delta z \rightarrow f^* \sim \delta z^{\frac{1}{1+\theta}}.
\end{align}
When the $N\delta z$ number of the weakest contacts are removed, we have
\begin{align}
 \abs{F_i} &\sim  \abs{-\frac{1}{\sqrt{N}}\sum_{\alpha=1}^{N\delta z}
 (P\bxi^{\mu_\alpha})_if_{\mu_\alpha}}
 \sim \frac{f_{\rm av}}{\sqrt{N}}\left(N\delta z\right)^{1/2}\new
 &\sim f_*\delta z^{1/2},\label{160859_8Aug19}
\end{align}
where $\mu_{\alpha}$ denotes the suffix of the weakest contacts. Here we
have assumed that $f_{\mu_\alpha}$ is uncorrelated with
$(P\bxi^{\mu_\alpha})_i$ and replaced it by its average
value. Substituting Eq.~(\ref{160859_8Aug19}) into
Eq.~(\ref{115457_22Sep19}), we finally arrive to
\begin{align}
 \lambda_{1} 
 \sim f_*^2 \delta z \sim \delta z^{\frac{3+\theta}{1+\theta}}.\label{165431_9Aug19}
\end{align} 
Interestingly, despite the difference of the dynamics and model, the
same result was previously derived for spherical particles driven by
shear~\cite{lerner2012toward}.  Using the result of the static replica
calculation
$\theta=0.423$~\cite{franz2017universality,charbonneau2014fractal}, we
have a theoretical prediction for the dynamical critical exponent
\begin{align}
\beta_{\rm theory} = \frac{3+\theta}{1+\theta} = 2.41.\label{162306_23Sep19}
\end{align}
This is reasonably close to the numerical result
Eq.~(\ref{174716_20Sep19}).

In Eq.~(\ref{160859_8Aug19}), we used the central limit theorem to
replace the summation of the $N\delta z$ random variables by $(N\delta
z)^{1/2}$. This would be verified if $N\delta z\gg 1$. On the contrary,
if $N\delta z\approx 1$, Eq.~(\ref{160859_8Aug19}) and the scaling
Eq.~(\ref{165431_9Aug19}) do not hold. In other words, the finite $N$
effects appear at $\delta z\sim 1/N$, which supports the scaling form
Eq.~(\ref{174654_20Sep19}) used for the finite size scaling analysis.

For the suspension flow of particle systems in finite dimensions, 
another theory that predicts a larger value of the critical exponent than 
the one predicted by 
Eq.~(\ref{162306_23Sep19}) is proposed~\cite{degiuli2015unified}. 
Further studies are necessary to understand such difference in critical 
exponents between GDD of the current model and suspension flow of particle systems.

\subsection{Scaling of the second eigenvalue}
In the previous subsection, we have discussed that the unbalanced force
controls the first eigenmode in the satisfiable phase. At the transition
point, the unbalanced force vanishes, which yields a strong correlation
between $\bxi^\mu$ along the direction of the unbalanced force
$\bF$. For the directions orthogonal to $\bF$, there are no such
constraints. Thus, we can assume that $(P\xi^\mu)_i$ are uncorrelated
with each other. In this case, the Hessian, Eq.~(\ref{121817_1Aug19}),
can be identified by a Wishart
matrix~\cite{livan2018introduction}. The eigenvalue distribution
$\rho(\lambda)$ is given by the Marchenko-Pastur
distribution~\cite{franz2015universal}:
\begin{align}
 \rho(\lambda) = (1-z)\delta(\lambda) +
 \frac{1}{2\pi}\frac{\sqrt{(\lambda-\lambda_-)(\lambda_+-\lambda)}}{\lambda},\label{115649_23Sep19}
\end{align}
where
\begin{align}
 \lambda_{\pm} &= \left(\sqrt{z}\pm 1\right)^2.\label{115922_23Sep19}
\end{align}
We believe that it would correctly describe the continuous
part of the spectrum. In Fig.~\ref{151720_9Sep19}, we saw that the
second eigenvalue $\lambda_{2}$ is the lowest eigenvalue of the
continuous spectrum. Therefore, from Eqs.~(\ref{115649_23Sep19}) and
(\ref{115922_23Sep19}), we expect for $\delta z\ll 1$
\begin{align}
 \lambda_{2}\sim \lambda_{-}\sim \delta z^2.
\end{align}
This expression is valid in the thermodynamic limit.  For finite $N$, we
put a similar Ansatz as Eq.~(\ref{174654_20Sep19}):
\begin{align}
\lambda_{2}(N,\delta z)\sim N^{-2}f_2(N\delta z),\label{151500_23Sep19}
\end{align}
where $f_2(x)\sim x^2$ for $x\gg 1$, and $f_2(x)\sim x^0$ for $x\sim
1$. 
\begin{figure}[t]
\includegraphics[width=9cm]{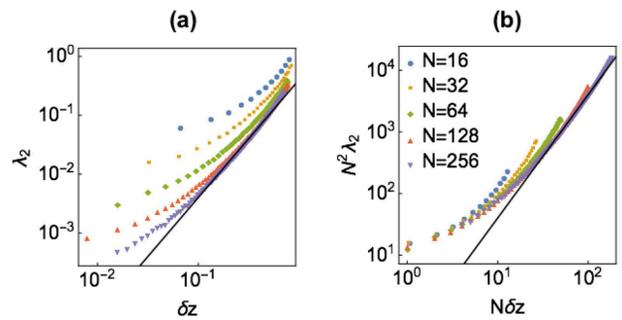} \caption{ Scaling of the
second eigenvalue $\lambda_2$ for $\sigma=-0.5$.  (a) The markers denote
 the numerical results of the average value of $\lambda_2$. The solid
lines denote the power law $\lambda_2\sim\delta z^2$.  (b) The scaling
plot of the same data.  } \label{151315_23Sep19}
\end{figure}
In Fig.~\ref{151315_23Sep19} (a), we show the numerical result of
$\lambda_2$ as a function of $\delta z$. One can see that $\lambda_2$
exhibits power law scaling $\lambda_2\sim \delta z^2$ for intermediate
values of $\delta z$. The power law region increases with $N$. In
Fig.~\ref{151315_23Sep19} (b), we show the scaling plot of the same
data. The collapse of the data for large $N$ and small $\delta z$
confirms Eq.~(\ref{151500_23Sep19}).

The above analysis shows that the first eigenmode $\lambda_1$ and the
continuous part of the spectrum are controlled by the completely
different mechanisms, which may explain the isolation of $\lambda_1$.

\section{Summary and discussions}
\label{164328_12Aug19}

In this work, we numerically studied the critical dynamics of the
perceptron near the SAT-UNSAT transition point.  The relaxation time is
inversely proportional to the first non-zero eigenvalue $\lambda_1$.  As
the system approaches the transition point, $\lambda_1$ vanishes much
faster than the continuous part of the spectrum. We discussed that
$\lambda_1$ is controlled by the unbalanced force which vanishes at the
transition point by construction. We then calculated the critical
exponent of $\lambda_1$ in the non-convex phase where the model has the
same universality as that of the spherical particles in the large
dimensional limit. We found $\lambda_1\sim \delta z^{2.55}$, which is
very close to the previous analytical result of frictionless spherical
particles driven by the external shear near the jamming transition
point~\cite{lerner2012toward}.

\begin{figure}[t]
\includegraphics[width=8cm]{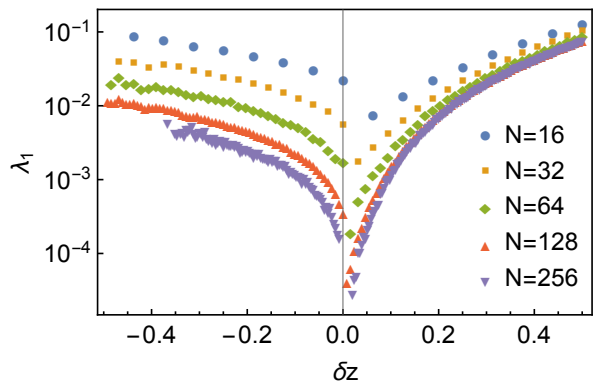} \caption{ $\delta z$
 dependence of the first eigenvalue $\lambda_1$ for $\sigma=-0.5$.
The markers denote the numerical results of the average value of
 $\lambda_1$. }
 \label{154805_30Sep19}
\end{figure}
One of our main findings is that the isolated mode robustly appears in
the SAT phase in the proximity of the SAT-UNSAT transition point both in
the case of convex and non-convex problems. This is a consequence of the
requirement of the force balance that yields non-trivial correlations
between the components of the Hessian.  As this is a quite general
mechanism for the models driven by the gradient descent dynamics, our
result raises a serious question about the usefulness of the
conventional stability analysis for complex systems based on random
matrix with uncorrelated
elements~\cite{may1972will,lecun1991second}. Further studies are
necessary along this line.

We find that the cost function exhibits exponential decay in the SAT
phase. In particular, the relaxation time remains finite even in the
non-convex phase. We would like to stress that this is qualitatively
different from the UNSAT phase.  To see this point more concretely, in
Fig.~\ref{154805_30Sep19}, we show the $\delta z=1-z$ dependence of the
first eigenvalue $\lambda_1$, which is inversely proportional to the
relaxation time.  One can see that in the SAT phase $\delta z>0$,
$\lambda_1$ converges to a finite value in the thermodynamic limit $N\to
\infty$ except very near the transition point $\delta z\ll 1$, in
particular, the data for $N=128$ and $N=256$ are almost
indistinguishable in the linear scale. On the contrary, in the UNSAT
phase $\delta z<0$, $\lambda_1$ exhibits significant finite size effects
even far from the transition point. This strong $N$ dependence in the
UNSAT phase is fully consistent with the previous theoretical result
based on the replica method, which predicts that the eigenvalue
distribution in the non-convex UNSAT phase is gapless in the
thermodynamic limit~\cite{franz2015universal}. This result gives some
theoretical background on the efficiency of the learning of neural
networks in the overparameterized region over that in the
underparameterized region.

Fig.~\ref{154805_30Sep19} shows that the minimum value of $\lambda_1$
shifts to rightward as $N$ decreases. This is a natural finite size effect as explained
below. $\lambda_1$ takes a minimal value when the system is isostatic:
the number of degrees of freedom $N-1$ is the same as that of
constraints $zN$, namely, $\delta z = 1/N$. This implies that the minimal
value of $\lambda_1$ shifts rightward in decreasing $N$.

The perceptron model investigated here belongs to the same universality
class of spherical particles in the large dimensional
limit~\cite{franz2016simplest}. From a practical point of view, it is
important to introduce the effect of asphericity, as real granular
particles are in general non-spherical. Moreover, there is a recent
study that reports some class of multilayer perceptron exhibits similar
empirical observations near the SAT-UNSAT phase transition to those of
the jamming of ellipsoids~\cite{Geiger2019}.  In previous works, we have
shown that the eigenvalue distribution of ellipsoids is significantly
different from that of spherical
particles~\cite{brito2018universality,ikeda2019mean,ikeda2019infinitesimal}. It
would be interesting to see how this difference affects the dynamics.

\begin{acknowledgments}
We thank F.~Zamponi, A.~Ikeda, E.~DeGiuli, A.~Altieri and P.~Urbani for
kind discussions. We also thank G.~Biroli for useful comments. This
project has received funding from the European Research Council (ERC)
under the European Union's Horizon 2020 research and innovation
programme (grant agreement n.~723955-GlassUniversality).
\end{acknowledgments}

\bibliography{apssamp}
\end{document}